\documentclass[12pt]{article}
\input{psfig.sty}
\input{epsf.sty}
\newcommand{\Define}{\stackrel{\triangle}{=}}

\begin{document}
\baselineskip 0.25in

\title{\LARGE Improved Linear Parallel Interference Cancellers\footnote{This
paper in part was presented in the IEEE International 
Conference on Communications (ICC'2007), Glasgow, June 2007, and in the
National Conference on Communications (NCC'2007), Indian Institute of
Technology, Kanpur, India. \newline
This work was supported in part by the Air Force Office
of Scientific Research under grant \# FA9550-06-1-0210.
}}
\author{T. Srikanth$^\dagger$, K. Vishnu Vardhan$^\dagger$, 
A. Chockalingam$^\dagger$, and L. B. Milstein$^\ddagger$
\\
{\normalsize $\dagger$ Department of ECE, Indian Institute of Science, 
Bangalore 560012, INDIA}  \\
{\normalsize $\ddagger$ Department of ECE, University of California,
San Diego, La Jolla 92093, USA}
}
\date{}
\maketitle
\vspace{-10mm}
\begin{center}
(Accepted in IEEE Trans. on Wireless Communications)
\end{center}
\baselineskip 2.00pc
\begin{abstract}
In this paper, taking the view that a linear parallel interference 
canceller (LPIC) can be seen as a linear matrix filter, we propose 
new linear matrix filters that can result in improved bit error
performance compared to other LPICs in the literature. The motivation 
for the proposed filters arises from the possibility of avoiding the 
generation of certain interference and noise terms in a given stage 
that would have been present in a conventional LPIC (CLPIC). In 
the proposed filters, we achieve such avoidance of the generation of 
interference and noise terms in a given stage by simply making the 
diagonal elements of a certain matrix in that stage equal to zero. 
Hence, the proposed filters do not require additional complexity 
compared to the CLPIC, and they can allow achieving a certain error 
performance using fewer LPIC stages. We also extend the proposed 
matrix filter solutions to a multicarrier DS-CDMA system, where we 
consider two types of receivers. In one receiver (referred to as 
Type-I receiver), LPIC is performed on each subcarrier first, 
followed by multicarrier combining (MCC). In the other receiver 
(called Type-II receiver), MCC is performed first, followed by LPIC. 
We show that in both Type-I and Type-II receivers, the proposed 
matrix filters outperform other matrix filters. Also, Type-II 
receiver performs better than Type-I receiver because of enhanced 
accuracy of the interference estimates achieved due to frequency 
diversity offered by MCC.
\end{abstract}

\vspace{-3mm}
\thispagestyle{empty}
\baselineskip 1.80pc

\newpage

\section{Introduction}
\label{sec1}
\vspace{-6mm}
Linear parallel interference cancellers (LPIC) have the
advantages of implementation simplicity, analytical tractability, and
good performance \cite{DSR98}-\cite{icc05}. The conventional way to
realize LPIC schemes is to use unscaled values of the previous stage
soft outputs of different users for multiple access interference (MAI)
estimation. In \cite{GRL00}, Guo {\em et al} described and analyzed
LPIC schemes for CDMA using a matrix-algebraic approach. They pointed
out that an LPIC can be viewed as a {\em linear matrix filter} applied
directly to the chip matched filter (MF) output vector. While the matrix
filter corresponding to the conventional LPIC (CLPIC) converges to the
decorrelating (DC) detector, they also proposed a modified matrix filter
which converges to a minimum mean square error (MMSE) detector. This was
done by exploiting the equivalence of the LPIC to a steepest descent 
optimization method for minimizing the mean square error. For this 
optimization, they obtained optimum step sizes for different stages 
that remove the excess mean square error in $K$ stages (where $K$ is 
the number of users), leaving only the minimum MSE in stages greater 
than $K$. The condition for this convergence has been shown to be that 
the maximum eigenvalue of the correlation matrix must be less than two.

Our contribution in this paper is that we propose new linear matrix
filters that can perform better than the matrix filters studied in
\cite{GRL00}. The motivation for the proposed filters arises from the
possibility of avoiding the generation of certain interference and
noise terms in a given stage that would have been present in the CLPIC.
In the proposed filters, we achieve such avoidance of the generation
of interference and noise terms in a given stage by simply making the
diagonal elements of a certain matrix in that stage equal to zero. Hence 
the proposed filters do not require additional complexity compared to the
CLPIC. We show that the proposed matrix filters can achieve better 
performance compared to other matrix filters in the literature. This,
in turn, can allow achieving a certain error performance using fewer
LPIC stages. We also propose filters that use different step sizes for
different stages (but the same step size for all users at a given stage).
In addition, we propose filters that use different weights for different
users in different stages, where we also obtain closed-form expressions 
for the optimum weights that maximize the output average SINR in 
a given stage.

We further extend the proposed matrix filter solutions to a multicarrier 
DS-CDMA system, where multicarrier combining (MCC) needs to be carried 
out in addition to the LPIC operation. Depending on which operation
(i.e., MCC or LPIC) gets done first at the receiver, the resulting 
performances and complexities differ. We consider two types of receivers. 
In what we call the Type-I receiver, LPIC is performed on each subcarrier 
first, followed by MCC \cite{icc05}. In the Type-II receiver, MCC is 
performed first, followed by LPIC. We show that in both Type-I and 
Type-II receivers, the proposed matrix filters outperform other matrix 
filters. Also, Type-II receiver outperforms Type-I receiver because of 
enhanced accuracy of the MAI estimates achieved due to frequency 
diversity offered by MCC.

The rest of the paper is organized as follows.  In Sec. \ref{sec2}, we
present the system model. In Sec. \ref{sec3}, we present the proposed 
matrix filters for single carrier DS-CDMA, along with their bit error 
performance results. Section \ref{sec4} presents the proposed filters 
and their performance in multicarrier DS-CDMA. Conclusions are presented 
in Sec. \ref{sec5}.

\vspace{-2mm}
\section{System Model}
\label{sec2}
\vspace{-3mm}
We consider a $K$-user synchronous multicarrier DS-CDMA system with $M$ 
subcarriers. Let $b_k \in \{+1,-1\}$ denote the binary data symbol of the 
$k$th user, which is sent in parallel on $M$ subcarriers \cite{Hanzo1}. 
Let $P$ denote the number of chips-per-bit in the signature waveforms. 
It is assumed that the channel is frequency non-selective on each 
subcarrier and the fading is slow (assumed constant over one bit interval) 
and independent from one subcarrier to the other. We assume that the 
subcarriers are separated sufficiently apart so that inter-carrier 
interference is negligible.
                                                                                
Let ${\bf y}^{(1),(i)}=\left[y_1^{(1),(i)}\,\,y_2^{(1),(i)}\,\,\cdots\,\,y_K^{(1),(i)}\right]^T$
denote the $K$-length received signal vector\footnote{Vectors are denoted 
by boldface lowercase letters, and matrices are denoted by boldface 
uppercase letters. $[.]^T$ and $[.]^H$ denote transpose and conjugate 
transpose operations, respectively. $Re\{a\}$ and $Im\{a\}$ denote 
the real and imaginary parts of $a$.} at the MF output\footnote{We take
the MF output (i.e., the despread output) as the 1st stage output in the 
multistage LPIC receivers. 
So, the $^{(m),(i)}$ in the superscript of $y$ denotes the stage index $m$ 
and subcarrier index $i$.} on the $i${th} 
subcarrier; i.e., $y_k^{(1),(i)}$ is the MF output (i.e., 1st stage output)
of the $k$th user on the $i$th subcarrier, given by

\begin{equation}
y_k^{(1),(i)} = \underbrace{A_k b_k h_k^{(i)}}_{\footnotesize \mbox{desired signal}} + \underbrace{\sum_{j = 1,j\neq{k}}^{K} \rho_{kj}^{(i)}A_jb_jh_j^{(i)}}_{{\footnotesize \mbox{MAI}}} + \, \underbrace{n_k^{(i)}}_{\footnotesize \mbox{noise}}.
\label{3eq1x}
\end{equation}
The MF output vector ${\bf y}^{(1),(i)}$ can be written in the form
\begin{equation}
\label{eqnA}
{\bf y}^{(1),(i)} \, = \, {{\bf R}^{(i)}{\bf H}^{(i)}{\bf b}}+{\bf n}^{(i)},
\end{equation}
where ${\bf H}^{(i)}$ represents 
the {\normalsize $K\hspace{-1mm}\times
\hspace{-1mm}K$} channel matrix, given by
\begin{eqnarray}
{\bf H}^{(i)} & = & diag\left\{h_1^{(i)}, h_2^{(i)}, \cdots, h_K^{(i)} \right\},\label{eqnC}
\end{eqnarray}
and the channel coefficients $h_k^{(i)}$, $i=1,2,\cdots,M$, are
assumed to be i.i.d. complex Gaussian r.v's with zero mean and
$E\Big[\Big(Re\{h_{k}^{(i)}\}\Big)^2\Big]=E\Big[\Big(Im\{h_{k}^{(i)}\}\Big)^2\Big]=0.5$.
The matrix ${\bf R}^{(i)}$ is the $K\times K$ cross-correlation matrix on
the $i$th subcarrier, given by
\begin{eqnarray}
\label{eqnB}
{\bf R}^{(i)} & = & \left[\begin{array}{cccc}
        1 & \rho_{12}^{(i)} & \cdots &  \rho_{1K}^{(i)} \\
\rho_{21}^{(i)} & 1         & \cdots &  \rho_{2K}^{(i)} \\
       \vdots & \vdots & \ddots & \vdots \\
       \rho_{K1}^{(i)} & \rho_{K2}^{(i)} & \cdots & 1 \end{array}\right],
\end{eqnarray}
where $\rho_{lj}^{(i)} = \rho_{jl}^{(i)}$ is the normalized cross-correlation 
coefficient between the signature waveforms of the $l$th and $j$th users on 
the $i$th subcarrier. The $K$-length data vector ${\bf b}$ is given by
${\bf b} \, = \, \left[ \begin{array}{cccc}
     A_1b_1 & A_2b_2 & \cdots & A_Kb_K
     \end{array} \right]^T$,
where $A_k$ denotes the transmit amplitude of the $k$th user. 
The $K$-length noise vector ${\bf n}^{(i)}$ on the $i$th subcarrier is 
given by
${\bf n}^{(i)} \, = \, \left[\begin{array}{cccc}
n_1^{(i)} & n_2^{(i)} & \cdots & n_K^{(i)} \end{array}\right]^T$,
where $n_k^{(i)}$ denotes the additive noise component of the $k$th user
on the $i$th subcarrier,
which is assumed to be complex Gaussian with zero mean with
{\normalsize $E[n_k^{(i)} \big(n_j^{(i)}\big)^{*}]=\sigma^2$}
when $j=k$ and
{\normalsize $E[n_k^{(i)} \big(n_j^{(i)}\big)^{*}]=\sigma^2\rho_{kj}^{(i)}$}
when $j\neq k$.

\vspace{-2mm}
\section{Proposed Linear Matrix Filters}
\label{sec3}
\vspace{-3mm}
In this section, we propose improved LPICs for single carrier DS-CDMA 
(i.e., $M=1$ in the system model presented in Sec. \ref{sec2}). So, for 
notational simplicity, we drop the subcarrier index $(i)$ in this 
section. In Sec. \ref{sec4}, we will consider $M>1$. We assume that 
all the channel coefficients are perfectly known at the receiver. 
Dropping the subcarrier index in (\ref{3eq1x}), the MF output 
of the desired user $k$, $y_k^{(1)}$, can be written as 

\begin{equation}
y_k^{(1)} = \underbrace{x_k}_{\footnotesize \mbox{desired signal}} + \underbrace{\sum_{j = 1,j\neq{k}}^{K} x_j \rho_{kj}}_{{\footnotesize \mbox{MAI}}} + \, \underbrace{n_k}_{\footnotesize \mbox{noise}}, \mbox{\hspace{5mm} where \,\,}
x_l \Define A_l b_l h_l, \,\,\,\,\, l=1,2,\cdots,K.
\label{3eq1}
\end{equation}

\vspace{-5mm}
\subsection{Conventional Matrix Filter, ${\bf G}^{(m)}$}
\label{sec21a}
\vspace{-3mm}
In conventional LPIC (CLPIC), an estimate of the MAI for the desired user
in the current stage is obtained using all the other users' soft outputs
from the previous stage for cancellation in the current stage. The $m$th 
stage output of the desired user $k$, $y_k^{(m)}$, in CLPIC is 
\cite{TMC06}
\vspace{-3mm}
\begin{eqnarray}
\label{PICNc}
y_k^{(m)} & = & y_k^{(1)} - \underbrace{\sum_{j=1, \, j\neq{k}}^{K} {\rho_{jk}y_j^{(m-1)}}}_{\mbox{\footnotesize MAI estimate}}.
\end{eqnarray}
The $k$th user's bit decision after MAI cancellation in the $m$th stage, 
$\widehat{b}^{(m)}_k$, is obtained as
\begin{eqnarray}
\widehat{b}^{(m)}_k = \mbox{sgn}\left(\mbox{Re}\Big(h_k^* y_k^{(m)}
\Big)\right).
\label{C}
\end{eqnarray}
The CLPIC output in (\ref{PICNc}) can be written in matrix algebraic
form as \cite{GRL00}
\vspace{-2mm}
\begin{eqnarray}
\mbox{\hspace{-6mm}}
{\bf y}^{(m)} & \mbox{\hspace{-4mm}} = \mbox{\hspace{-2mm}} & \mbox{\hspace{-2mm}} \left[{\bf I}+({\bf I}-{\bf R})+({\bf I}-{\bf R})^2+\cdots+({\bf I}-{\bf R})^{m-1}\right]{\bf y}^{(1)} \label{lf1} 
\,\, = \,\, \underbrace{\sum_{j=1}^{m}({\bf I}-{\bf R})^{j-1}}_{{\bf G}^{(m)}} \,\,\, {\bf y}^{(1)}, 
\label{lf2}
\end{eqnarray}
The ${\bf G}^{(m)}$ filter in (\ref{lf2}) can be viewed as an equivalent
one-shot linear matrix filter for the $m$th stage of the CLPIC.

\vspace{-3mm}
\subsection{Proposed Matrix Filter, ${\bf G_p}^{(m)}$}
\label{sec31}
\vspace{-3mm}
In this subsection, we propose a new linear matrix filter, which we denote
as ${\bf G_p}^{(m)}$, that can perform better than the matrix filter 
${\bf G}^{(m)}$ in (\ref{lf2}). The motivation for the new matrix filter 
can be explained as follows.

{{\bf What does the matrix filter ${\bf G}^{(m)}$ do:}
It is noted that the behavior of the ${\bf G}^{(m)}$ filter in 
(\ref{lf2}) (i.e., CLPIC) at a given stage $m\geq 2$ is characterized 
by $a)$ interference removal, $b)$ generation of new interference terms, 
$c)$ desired signal loss/gain, $d)$ desired signal recovery/removal, 
and $e$) noise enhancement. For example, the cancellation operation 
in the 2nd stage (i.e., $m=2$) results in $i)$ interference removal, 
$ii)$ generation of new interference terms, $iii)$ desired signal 
loss, and $iv)$ noise enhancement. This can be seen by observing the 
2nd stage output expression for the desired user $k$, which can be 
written, using (\ref{PICNc}) and (\ref{3eq1}), as
\begin{eqnarray}
y_k^{(2)} & = & y_k^{(1)} - \sum_{j=1, \, j\neq{k}}^{K} {\rho_{jk}y_j^{(1)}} \nonumber \\
& = & \left(x_k + \sum_{i=1,i\neq k}^{K}\rho_{ki} x_i + n_k\right) 
 - \sum_{j=1,j\neq k}^{K} \rho_{jk} \left(x_j + \sum_{\underbrace{l=1, l\neq j}_{\mbox{\footnotesize{$l$ can be $k$ here}}}}^{K}\rho_{jl}x_l +n_j  \right)\nonumber \\
& = & x_k - \underbrace{x_k\sum_{j=1,j\neq k}^{K}\rho_{jk}^2}_{\mbox{{\footnotesize desired signal loss}}} 
 - \underbrace{\sum_{j=1, j\neq k}^{K} \rho_{jk} \sum_{l=1, l\neq j,k}^{K} \rho_{jl} x_l}_{\mbox{{\footnotesize new interference terms}}} 
+ \,\,\,\, n_k - \underbrace{\mbox{\hspace{-2mm}} \sum_{j=1, j\neq k}^{K} \rho_{jk} n_j}_{\mbox{{\footnotesize additional noise terms}}}\mbox{\hspace{-5mm}},
\label{A2}
\end{eqnarray}
Comparing the expression at the MF output, $y_k^{(1)}$, in (\ref{3eq1})
and the expression for the
2nd stage output, $y_k^{(2)}$, in (\ref{A2}), it can be seen that
the cancellation operation in the 2nd stage results in the following
at the 2nd stage output.
\vspace{-6mm}
\begin{itemize}
\item   The interference terms, $\sum_{j\neq{k}}^{} \rho_{jk}x_j$,
        that were present in the MF output in (\ref{3eq1}) are removed.
	In the process, $i$) new interference terms proportional to 
	$\rho^2$, i.e., 
	$\sum_{j\neq k}^{} \rho_{jk} \sum_{l\neq j,k}^{} \rho_{jl} x_l$
	in (\ref{A2}), get generated, 
	$ii$) a fraction $\sum_{j\neq k}^{}\rho_{jk}^2$ of the desired 
	signal component gets lost, and 
	$iii$) additional noise terms proportional to $\rho$, i.e.,
	$\sum_{j\neq k}^{} \rho_{jk} n_j$ in (\ref{A2}), get 
	introduced. 
\end{itemize}
\vspace{-5mm}
In Appendix A, we present the expression for the 3rd stage output in 
an expanded form. From (\ref{3rd}) in Appendix A, we can 
make the following observations which result from the cancellation operation 
in the 3rd stage. 
\vspace{-5mm}
\begin{itemize}
\item   The desired signal loss that occurred in the 2nd stage is recovered
	(see the two {\tiny \fbox{$A$}} terms cancelling each other in 
	(\ref{3rd})). In the process, new interference terms proportional 
	to $\rho^3$ (see the {\tiny \fbox{$B_I$}} term in (\ref{3rd})) as 
	well as additional noise terms proportional to $\rho^2$ (see the 
	{\tiny \fbox{$B_N$}} term in (\ref{3rd})) get generated.
\item   Interference terms generated in the 2nd stage are removed (see the 
	two {\tiny \fbox{$C$}} terms cancelling each other in (\ref{3rd})). 
	In the process, $i$) further desired signal loss/gain\footnote{Depending	on $\rho$'s being positive or negative, the term {\scriptsize 
	\fbox{$D$}} in (\ref{3rd}) can be positive or negative, because of 
	which 
	there can be a desired signal gain or loss.} proportional 
	to $\rho^3$ occurs (see the {\scriptsize \fbox{$D$}} term in 
	(\ref{3rd})), and $ii$) new interference terms proportional
        to $\rho^3$ (see the {\tiny \fbox{$E_I$}} term in (\ref{3rd})) as
        well as additional noise terms proportional to $\rho^2$ (see the
        {\tiny \fbox{$E_N$}} term in (\ref{3rd})) get generated.
\end{itemize}
\vspace{-4mm}
Similar observations can be made on the expanded form of the equations
for the subsequent stages of the CLPIC\footnote{The general expression 
for the $m$th stage output in expanded form, for any $m \geq 3$, and 
the corresponding observations are given in Appendix B.}.
For $m \rightarrow \infty$, the CLPIC is known to converge to the 
decorrelating detector, provided the eigenvalues of the ${\bf R}$ 
matrix are less than two \cite{GRL00}. That is, when 
$m\rightarrow \infty$, in the expression for ${\bf G}^{(m)}$ output
in (\ref{3eq101}), the desired signal loss/gain and the interference 
terms go to zero and the noise term gets enhanced.

{\bf What is proposed to be achieved using the ${\bf G_p}^{(m)}$:} 
As explained above, in the ${\bf G}^{(m)}$ filter, new interference 
and noise terms get generated in the process of interference removal 
and recovery/removal of desired signal loss/gain. We seek to avoid the
generation of some of these new interference and noise terms. For example, 
as will be shown next, the generation of the {\tiny \fbox{$B_I$}} and 
{\tiny \fbox{$B_N$}} terms at the 3rd stage output in (\ref{3rd}) can be 
avoided by simply making the diagonal elements of a certain matrix in the 
cancellation operation in the 3rd stage equal to zero. This, as we will 
see later, can result in improved performance compared to the 
${\bf G}^{(m)}$ filter.

{\bf Proposed matrix filter, ${\bf G_p}^{(m)}$:}
We propose to avoid the generation of new interference and noise terms 
in $T_3$ in (\ref{3eq91}), caused in the process of recovery/removal 
of desired signal loss/gain in the previous stage. Since there is no 
desired signal loss/gain in the 1st stage, the 2nd stage of the proposed
filter is the same as that of the ${\bf G}^{(m)}$ filter, i.e.,
${\bf G_p}^{(2)} = {\bf G}^{(2)}$. For stages greater than two, i.e., 
for $m \geq 3$, the $m$th stage output of the proposed filter 
${\bf G_p}^{(m)}$, denoted by $y_{k,{{\tiny {\bf p}}}}^{(m)}$, can be 
written as
\begin{eqnarray}
y_{k,{{\tiny {\bf p}}}}^{(m)} & = & y_{k,{{\tiny {\bf p}}}}^{(m-1)} 
+ \, (-1)^{m+1}\hspace{-0mm}\sum_{k_1\ne k}^{K} \, \sum_{k_2\ne k,k_1}^{K}\hspace{-0mm}\, \sum_{k_3\ne k,k_2}^{K}\cdots  \nonumber \\
& & \mbox{\hspace{-0mm}} 
\sum_{k_{m-2}\ne k,k_{m-3}}^{K} \, \sum_{k_{m-1}\ne k,k_{m-2}}^{K}\hspace{-6mm}\rho_{kk_{m-1}}\rho_{k_{m-1}k_{m-2}}\cdots
\rho_{k_3k_2}\rho_{k_2k_1} \, y_{k_1}^{(1)}.
\label{eqx9}
\end{eqnarray}
We note that the above expression is obtained by $i$) dropping $T_3$ from 
(\ref{3eq91}), and $ii$) modifying $T_4$ in (\ref{3eq91}) such that all 
the summations in it exclude the desired user index $k$. The above two 
modifications ensure that the proposed filter removes the previous stage 
interference while avoiding the recovery/removal of the desired signal 
loss/gain\footnote{Although possible signal loss recovery is avoided in 
the process, the net effect can still be beneficial (we will see this 
in Sec. \ref{sec51}).}. Also, because of these modifications, the 
interference and noise terms in a given stage of the proposed filter 
will be a subset of the interference and noise terms in the same stage 
of the ${\bf G}^{(m)}$ filter. Equation (\ref{eqx9}) can be written in 
the following form
\begin{eqnarray}
\mbox{\hspace{-14mm}}
y_{k,{\tiny {\bf p}}}^{(m)} & = & y_k^{(1)} - \sum_{k_1\ne k}^{K} \Bigg(\rho_{kk_1}-\sum_{k_2\ne k,k_1}^{K}\rho_{kk_2}\rho_{k_2k_1} 
+ \,\, \sum_{k_2\ne k,k_1}^{K}\sum_{k_3\ne k,k_2}^{K}\rho_{kk_3}\rho_{k_3k_2}\rho_{k_2k_1} - \,\, \cdots \nonumber \\
& & \mbox{\hspace{-18mm}} + \, (-1)^{m} \sum_{k_2\ne k,k_1}^{K} \, \sum_{k_3\ne k,k_2}^{K}\cdots \sum_{k_{m-2}\ne k,k_{m-3}}^{K} \,\,\sum_{k_{m-1}\ne k,k_{m-2}}^{K} 
\mbox{\hspace{-0mm}} \rho_{kk_{m-1}}\rho_{k_{m-1}k_{m-2}}\cdots\rho_{k_3k_2}\rho_{k_2k_1} \Bigg) \, y_{k_1}^{(1)},
\label{gx}
\end{eqnarray} 
which, in turn, can be expressed in matrix form as 
\begin{eqnarray}
{\bf y}_{\tiny {\bf p}}^{(m)} & = & \underbrace{\left(\sum_{j=0}^{m-1} {\bf B}_j\right)}_{{\footnotesize {\bf G_p}^{(m)}}} \,\, {\bf y}^{(1)},
\label{gp1}
\end{eqnarray}
where
\vspace{-4mm}
\begin{eqnarray}
{\bf B}_n & = & \Big[{\bf B}_{n-1}\left({\bf I}-{\bf R}\right)\Big]^{\odot},
\label{p2}
\end{eqnarray}
$[{\bf M}]^\odot$ denotes the matrix ${\bf M}$ with its diagonal
elements made equal to zero, and ${\bf B}_0 = {\bf I}$. 
Note that, since (\ref{gp1}) is structurally the same as (\ref{lf1}) and 
the $[.]^\odot$ operation in (\ref{p2}) does not add to complexity,
the proposed ${\bf G_p}^{(m)}$ filter has the same complexity as the
${\bf G}^{(m)}$ filter.

The ${\bf G}^{(m)}$ filter is known to converge 
to the decorrelating detector for $m \rightarrow \infty$, provided the 
maximum eigenvalue of the ${\bf R}$ matrix is less than two \cite{GRL00}. 
That is, ${\bf G}^{(\infty)}={\bf R}^{-1}$, which results in the output 
vector
\begin{eqnarray}
\left({\bf y}^{(\infty)}\right)_{\bf G} & = & {\bf R}^{-1}{\bf y}^{(1)} \,\,\,\, = \,\,\,\, {\bf x}+{\bf R}^{-1}{\bf n}.
\label{f1}
\end{eqnarray}
As with ${\bf G}^{(m)}$, all the interference terms in ${\bf G_p}^{(m)}$ 
also go to zero for $m\rightarrow \infty$. This can be seen as follows. 
From (\ref{gp1}) and (\ref{p2}), ${\bf G_p}^{(\infty)}$ can be written 
in the form 
\begin{eqnarray}
{\bf G_p}^{(\infty)} & = & 
\underbrace{{\bf I}}_{{\bf B}_0} + \, \underbrace{\left[({\bf I-R})-{\bf D}_1\right]}_{{\bf B}_1} +\, \underbrace{\left\{\left[({\bf I-R})-{\bf D}_1\right]({\bf I-R})-{\bf D}_2\right\}}_{{\bf B}_2} + \cdots, 
\label{gp1x}
\end{eqnarray}
where ${\bf D}_n$ is a diagonal matrix with the diagonal elements 
the same as those in the matrix ${\bf B}_{n-1}({\bf I-R})$. 
Equation (\ref{gp1x}) can be written as
\begin{eqnarray}
{\bf G_p}^{(\infty)} & = & \underbrace{\left({\bf I} + ({\bf I-R}) + ({\bf I-R})^2 + \cdots\right)}_{{\bf R}^{-1}} - \,\, {\bf D}_1 \left({\bf I} + ({\bf I-R}) + ({\bf I-R})^2 + \cdots\right) \nonumber \\
&  &  - \,\, {\bf D}_2 \left({\bf I} + ({\bf I-R}) + ({\bf I-R})^2 + \cdots\right) - \,\, \cdots \nonumber \\
& = &  \underbrace{\left({\bf I} - {\bf D}_1 - {\bf D}_2 - \cdots \right)}_{\Define \, {\bf F}} \,\, {\bf R}^{-1}. 
\label{gp2x}
\end{eqnarray}
Hence, the output vector for $m\rightarrow \infty$ is given by
\begin{eqnarray}
\left({\bf y}^{(\infty)}\right)_{\bf G_p} & = & {\bf F}{\bf R}^{-1}{\bf y}^{(1)} \,\,\,\, = \,\,\,\, {\bf F}{\bf x}+{\bf F}{\bf R}^{-1}{\bf n}.
\label{f2}
\end{eqnarray}
The diagonal matrix ${\bf F}$ defined in (\ref{gp2x}) can be written as  
\begin{eqnarray}
{\bf F} & = & diag\Big(f_1, f_2, \cdots, f_K\Big),
\end{eqnarray}
where $f_k$ is given by 
\begin{eqnarray}
\mbox{\hspace{-9mm}}
f_k & \mbox{\hspace{-2mm}} = & \mbox{\hspace{-2mm}} 1-\sum_{k_1 \neq k}^{K}\rho_{kk_1}\rho_{k_1k} + \sum_{k_1\neq k}^{K}\sum_{k_2\neq k,k_1}^{K} \mbox{\hspace{-2mm}} \rho_{kk_2}\rho_{k_2k_1}\rho_{k_1k} 
- \sum_{k_1\neq k}^{K}\, \sum_{k_2\neq k,k_1}^{K} \, \sum_{k_3 \neq k,k_2}^{K} \mbox{\hspace{-2mm}} \rho_{kk_3}\rho_{k_3k_2}\rho_{k_2k_1}\rho_{k_1k} \cdots 
\label{fk1}
\end{eqnarray}
For the case of equi-correlated users, $f_k$ in (\ref{fk1}) can be shown 
to converge to \\
${\small 1- \left((K-1)\rho^2/(1+(K-2)\rho \right)}$, and 
there are no interference terms in (\ref{f2}). Also, note that the outputs 
of the ${\bf G}$ filter in (\ref{f1}) and the ${\bf G_p}$ filter in 
(\ref{f2}) have the same SNR for $m\rightarrow \infty$. 

\vspace{-1mm}
\subsection{Why {\small ${\bf G_p}^{(m)}$} can perform better than 
{\small ${\bf G}^{(m)}$} - An Illustration}
\vspace{-2mm}
To analytically see why ${\bf G_p}^{(m)}$ can perform better than 
${\bf G}^{(m)}$ for $m\geq 3$, consider the case of $K$ equi-correlated 
users with correlation coefficient $\rho$, and no noise. Let us consider 
the average signal-to-interference ratio (SIR) at the 3rd stage output 
for ${\bf G_p}^{(m)}$ and ${\bf G}^{(m)}$.
The 3rd stage output of ${\bf G}^{(m)}$, in the absence of
noise, can be written as
\begin{eqnarray}
\left(y_k^{(3)}\right)_{{\bf G}} & = & \Big[1 + {\scriptsize \fbox{$D$}} \Big] x_k \, + \, {\scriptsize \fbox{$B_I$}} \, + \, {\scriptsize \fbox{$E_I$}}\,,
\label{gx3}
\end{eqnarray}
where the terms {\scriptsize \fbox{$D$}}\hspace{0.5mm},  
{\scriptsize \fbox{$B_I$}}\hspace{0.5mm},  
{\scriptsize \fbox{$E_I$}}\hspace{0.5mm} are defined in (\ref{3rd}).
Likewise, the 3rd stage output of ${\bf G_p}^{(m)}$ can be written as
\begin{eqnarray}
\left(y_k^{(3)}\right)_{{\bf G_p}} & = & \Big[1 - {\scriptsize \fbox{$A$}} + {\scriptsize \fbox{$D$}} \Big] x_k \, + \, {\scriptsize \fbox{$E_I$}}\,.
\label{gpx3}
\end{eqnarray}
Note that the interference term {\scriptsize \fbox{$B_I$}} generated
in ${\bf G}^{(m)}$ is not generated in ${\bf G_p}^{(m)}$. Also, the 
desired signal term {\scriptsize \fbox{$A$}} is recovered in ${\bf G}^{(m)}$ 
whereas it is not recovered in ${\bf G_p}^{(m)}$.
Now, from (\ref{gx3}), the average SIR at the 3rd stage output of 
${\bf G}^{(m)}$, for the case of equi-correlated and equal-amplitude 
users, can be obtained as
\begin{eqnarray}
\left(\overline{\mbox{SIR}}\right)_{{\bf G}}^{(3)}=\frac{\Big(1+(K-1)(K-2)\rho^3\Big)^2}{(K-1)\Big((K-1)\rho^3+(K-2)^2\rho^3\Big)^2}.
\label{nr}
\end{eqnarray}
Likewise, from (\ref{gpx3}), the average SIR at the 3rd stage output
of ${\bf G_p}^{(m)}$ can be obtained as
\begin{eqnarray}
\mbox{\hspace{-6mm}}
\left(\overline{\mbox{SIR}}\right)_{{\bf G_p}}^{(3)} & = &
\frac{\Big(1-(K-1)\rho^2+(K-1)(K-2)\rho^3\Big)^2}{(K-1)\Big((K-2)^2\rho^3\Big)^2}.
\label{dr}
\end{eqnarray}
From (\ref{nr}) and (\ref{dr}), it can be seen that
\begin{eqnarray}
\beta & \Define & \sqrt{\frac{\left(\overline{\mbox{SIR}}\right)_{{\bf G_p}}^{(3)}}{\left(\overline{\mbox{SIR}}\right)_{{\bf G}}^{(3)}}} 
\,\, = \,\, 1+\frac{(K-1)\Big(1-(K-2)\rho\Big)\Big(1+(K-2)\rho-(K-1)\rho^2\Big)}{(K-2)^2\Big(1+(K-1)(K-2)\rho^3\Big)}.
\label{a_val}
\end{eqnarray}
For $\rho>0$, the maximum eigenvalue for the ${\bf R}$ matrix is 
$1+(K-1)\rho$, so that the condition for convergence is 
$1+(K-1)\rho < 2$, i.e., $(K-1)\rho < 1$. Now, in (\ref{a_val}), 
the $2$nd term is positive when $(K-1)\rho < 1$, which results 
in $\beta > 1$. This implies that ${\bf G_p}^{(m)}$ results in a higher 
average output SIR than ${\bf G}^{(m)}$. As will be seen in Sec. 
\ref{sec51}, simulation results show that ${\bf G_p}^{(m)}$ can 
perform better than ${\bf G}^{(m)}$ in the case of non-equicorrelated 
users as well as in the presence of noise. 

\vspace{-3mm}
\subsection{A Modified MMSE Converging Filter, {\small ${\bf G_{p\mu}}^{(m)}$}}
\vspace{-2mm}
As pointed out in Sec. \ref{sec1}, Guo {\em et al}, in \cite{GRL00}, have 
proposed modifications to the ${\bf G}^{(m)}$ filter so that the resulting 
modified matrix filter converges to the MMSE detector instead of the
decorrelating detector, by exploiting the equivalence of the LPIC to the
steepest descent method (SDM) of optimization for minimizing the MSE. 
They also derived optimum step sizes for various stages, which ensured 
convergence to the MMSE detector in $K$ stages, where $K$ is the number of 
users. We refer to this MMSE converging matrix filter proposed by Guo 
{\em et al} in \cite{GRL00} as the ${\bf G_\mu}^{(m)}$ filter, which is 
given by \cite{GRL00} 

\begin{eqnarray}
{\bf y}_{\mu}^{(m)} & \mbox{\hspace{-4mm}} = & \mbox{\hspace{-4mm}} \underbrace{\left(\mu_m{\bf I}+\sum_{i=1}^{m-1}\mu_{m-i}\prod_{j=1}^{i}({\bf I}-\mu_{m-i+j}({\bf R}+\sigma^2{\bf I})) \right)}_{{\bf G_\mu}^{(m)}}{\bf y}^{(1)},
\label{gmu}
\end{eqnarray}
where $\mu_i$ is the step size at stage $i$, the optimum values of which
were obtained to be
\begin{eqnarray}
\mu_i=\frac{1}{\lambda_i+\sigma^2}, \,\,\,\, i=1,2,\cdots,K,
\label{eig}
\end{eqnarray}
where $\lambda_i, \,\, i=1,2,\cdots,K$ are eigenvalues of matrix ${\bf R}$. 
We note that a similar SDM view can be taken to modify our proposed
matrix filter ${\bf G_p}^{(m)}$ so that it can converge to the MMSE 
detector. We refer to such a modified version of our proposed filter 
as ${\bf G_{p\mu}}^{(m)}$ filter, where we avoid the generation of 
new interference and noise terms as in ${\bf G_{p}}^{(m)}$, while 
using the step sizes obtained for ${\bf G_\mu}^{(m)}$ in \cite{GRL00}. 
Accordingly, we propose the modified version of the MMSE converging 
filter as   
\begin{eqnarray}
{\bf y}_{{\tiny {\bf p}}{\mu}}^{(m)} & = & \underbrace{\left(\mu_m{\bf I}+\sum_{i=1}^{m-1}\mu_{m-i}{\bf J}_i \right)}_{{\bf G_{p\mu}}^{(m)}} {\bf y}^{(1)},
\label{mx1}
\end{eqnarray}
where ${\bf J}_i$ is given by
\begin{eqnarray}
{\bf J}_i & = & \left[{\bf J}_{i-1}\left({\bf I}-\mu_{K-i+1}({\bf R}+\sigma^2{\bf I})\right)\right]^{\odot}, \,\,\,\, \mbox{and} \,\,\,{\bf J}_0 = {\bf I}.
\end{eqnarray}

\vspace{-4mm}
\subsection{A Weighted Matrix Filter, 
${\bf G_{pw}}^{\mbox{\hspace{-2mm}}(m)}$}
\vspace{-3mm}
In the ${\bf G_{\mu}}^{(m)}$ and ${\bf G_{p\mu}}^{(m)}$ filters above, 
different step sizes are used in different stages (but the same step 
size for all users in a stage). Improved performance can be achieved if 
different scaling factors (weights) are used for different users in 
different stages. Accordingly, we propose a weighted version of our 
proposed filter ${\bf G_{p}}^{(m)}$. We refer to this weighted version 
as ${\bf G_{pw}}^{(m)}$, and is derived as follows.

In a weighted LPIC (WLPIC), the MAI estimate in a given stage is scaled
by a weight before cancellation (unit weight corresponds to CLPIC and
zero weight corresponds to MF). For example, the $m$th stage output of
the desired user $k$, $y_{k,{\tiny {\bf w}}}^{(m)}$, in a WLPIC is given by
\begin{eqnarray}
y_{k,{\tiny \bf w}}^{(m)} & = & y_k^{(1)} - w_k^{(m)} \sum_{j=1, \, j\neq{k}}^{K} {\rho_{jk}y_{j,{\tiny \bf w}}^{(m-1)}},
\label{PICN}
\end{eqnarray}
where $w_k^{(m)}$ is the weight with which the MAI estimate for the
$k$th user in the $m$th stage is scaled. For $m\geq 2$, the weighted 
cancellation operation in (\ref{PICN}) can be written in the form
\begin{eqnarray}
{\bf y}_{\tiny \bf w}^{(m)}& \mbox{\hspace{-3mm}} =& \mbox{\hspace{-2mm}} \Bigg({\bf I}+{\bf W}^{(m)}({\bf I}-{\bf R})+{\bf W}^{(m)}({\bf
I}-{\bf R}){\bf W}^{(m-1)}({\bf I}-{\bf R})+\cdots  \nonumber\\
& & \mbox{\hspace{0mm}} 
+ \, {\bf W}^{(m)}({\bf I}-{\bf R}){\bf W}^{(m-1)}({\bf I}-{\bf R})\cdots {\bf W}^{(2)}({\bf I}-{\bf R})\Bigg) \, {\bf y}^{(1)},
\label{stagem_w}
\end{eqnarray}
where ${\bf W}^{(m)}$ is the weight matrix at the $m$th stage, given by
${\small {\bf W}^{(m)} = \mbox {diag}\left(w_1^{(m)}, w_2^{(m)}, \cdots, w_K^{(m)}\right)}$,
and ${\bf W}^{(1)} = {\bf 0}$.
Now, as in ${\bf G}^{(m)}$, in order to avoid the generation of new 
interference and noise terms, we modify (\ref{stagem_w}) as follows:
\begin{eqnarray}
{\bf y}_{\tiny \bf pw}^{(m)} & = & \underbrace{\left(\sum_{j=0}^{m-1}\tilde{{\bf B}}_j\right)}_{{\bf G_{pw}}^{(m)}} \,\, {\bf y}^{(1)},
\label{gpw}
\end{eqnarray}
where
\vspace{-4mm}
\begin{eqnarray}
\tilde{{\bf B}}_n & = & \left[(\tilde{{\bf B}}_{n-1})({\bf W}^{(m-n+1)})({\bf I}-{\bf R})\right]^{\odot}, \,\,\, \mbox{and} \,\,\, \tilde{{\bf B}}_0={\bf I}.
\end{eqnarray}
Note that
${\bf G_{pw}}^{(m)}$ becomes ${\bf G_{p}}^{(m)}$ when
${\bf W}^{(m)} = {\bf I}, \,\, \forall \, m >1$. 

{\bf Optimum Weight Matrix, ${\bf W}^{(m)}_{\mbox{{\scriptsize opt}}}$}:
The $m$th stage output of the $k$th user when ${\bf G_{pw}}^{(m)}$
filter is used can be written as

\vspace{-6mm}
{\footnotesize
\begin{eqnarray}
\mbox{\hspace{-8mm}}
\left(y_k^{(m)}\right)_{{\bf G_{pw}}} & = & y_k^{(1)} - w_k^{(m)}\sum_{i\ne k}^{K}q_{k,i}^{(m)} y_i^{(1)} \nonumber \\
& \mbox{\hspace{-41mm}} = & \mbox{\hspace{-22mm}} \underbrace{x_k\Bigg(1-w_k^{(m)}\sum_{i\ne k}^{K}q_{k,i}^{(m)}\rho_{ki}\Bigg)}_{\mbox{desired signal}} 
+ \,\, \underbrace{n_k - w_k^{(m)}\sum_{i\ne k}^{K}q_{k,i}^{(m)}n_i}_{\mbox{{\footnotesize noise}}} 
+ \,\, \underbrace{\sum_{i\ne k}^{K}\Bigg(\rho_{ki}-w_k^{(m)}\Bigg(q_{k,i}^{(m)}+\sum_{k_1\neq i,k}^{K}q_{k,k_1}^{(m)}\rho_{k_1i}\Bigg)\Bigg)x_i}_{\mbox{{\footnotesize interference}}}, 
\label{nox}
\end{eqnarray}}

\vspace{-6mm}
where

\vspace{-8mm}
{\small
\begin{eqnarray}
\mbox{\hspace{-6mm}}
q_{k,i}^{(m)} & = & \Bigg(\rho_{ki}-\sum_{k_1\ne k,i}^{K}w_{k_1}^{(m-1)}\rho_{kk_1}\rho_{k_1i}+\sum_{k_1\ne k,i}^{K}w_{k_1}^{(m-2)}\sum_{k_2\ne k,k_1}^{K} 
\mbox{\hspace{-0mm}} w_{k_2}^{(m-1)}\rho_{kk_2}\rho_{k_2k_1}\rho_{k_1i} -\cdots \nonumber \\
& & \mbox{\hspace{-18mm}} + (-1)^m \mbox{\hspace{-1mm}} \sum_{k_1\ne k,i}^{K} \mbox{\hspace{-2mm}} w_{k_1}^{(2)}\sum_{k_2\ne k,k_1}^{K}\mbox{\hspace{-3mm}}w_{k_2}^{(3)} 
\mbox{\hspace{-0mm}} \cdots \,\mbox{\hspace{-5mm}} \sum_{k_{m-3}\ne k,k_{m-4}}^{K} \mbox{\hspace{-6mm}} w_{k_{m-3}}^{(m-2)}\mbox{\hspace{-2mm}}\sum_{k_{m-2}\ne k,k_{m-3}}^{K} \mbox{\hspace{-6mm}} w_{k_{m-2}}^{(m-1)} 
\mbox{\hspace{-0mm}} \rho_{kk_{m-2}}\rho_{k_{m-2}k_{m-3}}\cdots\rho_{k_3k_2}\rho_{k_2k_1}\rho_{k_1i} \Bigg). 
\end{eqnarray}}

\vspace{-10mm}
Since the interference and noise terms in on the RHS of (\ref{nox}) are
the sum of linear combinations of complex Gaussian r.v's (since the 
fade coefficients ${h_k}$ are assumed to be complex Gaussian), 
the average SINR for the $k$th user at the $m$th stage output can be 
obtained, in closed-form, as
\begin{eqnarray}
\overline{\mbox{SINR}}_k^{(m)} & = & \frac{A_k^2\Big(1-a w_k^{(m)}\Big)^2}{\sigma_I^2+\sigma_N^2},
\label{sinr}
\end{eqnarray}
where
\begin{eqnarray*}
a & = & \sum_{i\ne k}^{K}q_{k,i}^{(m)}\rho_{ki}, 
\,\,\,\,\,\,\,\,\,\,\,
b \,\, = \,\, \sum_{i\ne k}^{K}\rho_{ki}^2A_i^2,
\,\,\,\,\,\,\,\,\,\,\,
c \,\, = \,\, \sum_{i\ne k}^{K}\Big(q_{k,i}^{(m)}+\sum_{k_1\neq i,k}^{K}q_{k,k_1}^{(m)}\rho_{k_1i}\Big)^2A_i^2, \nonumber \\
d & = & \sum_{i\ne k}^{K}\rho_{ki}\Big(q_{k,i}^{(m)}+\sum_{k_1\neq i,k}^{K}q_{k,k_1}^{(m)}\rho_{k_1i}\Big)A_i^2,
\,\,\,\,\,\,\,\,\,\,\,
e \,\, = \,\,  \sum_{i\ne k}^{K}\sum_{j\ne k}^{K}q_{k,i}^{(m)}q_{k,j}^{(m)}\rho_{ij}, \nonumber \\
\sigma_I^2 & = & b+\left(w_k^{(m)}\right)^2 c - 2 \,w_k^{(m)} d,
\,\,\,\,\,\,\,\,\,\,\,
\sigma_N^2 \,\, = \,\, \sigma^2\left(1+\left(w_k^{(m)}\right)^2 e - 2 \, w_k^{(m)} a\right).
\end{eqnarray*}
By differentiating the average SINR expression in (\ref{sinr}) w.r.t.
$w_k^{(m)}$ and equating to zero, the optimum weights $w_{k,opt}^{(m)}$ 
can be obtained, in closed-form, as
\begin{eqnarray}
w_{k,opt}^{(m)} & = & \frac{d-ab}{c-ad+\sigma^2(e-a^2)}.
\label{opt}
\end{eqnarray}
In Fig. \ref{fig1}, we plot the average SINR at the $m$th stage output 
of the proposed ${\bf G_{pw}}^{(m)}$ filter, as a function of weight, 
$w_k^{(m)}$, at an average SNR of 20 dB. The average SNR of user $k$ is 
defined as $A_k^2/\sigma^2$. The number of users
considered is $K=20$, the processing gain is $P=64$, and there is no 
near-far effect (i.e., $A_1=A_2=\cdots=A_K$). From Fig. \ref{fig1},
it can be seen that for a given stage index $m$, the maximum
output average SINR occurs at an optimum weight; the closed-form
expression for this optimum weight is given by (\ref{opt}). The
maximum average SINR increases as $m$ is increased. Also, we see
diminishing improvement in SINR with increasing $m$, as expected. 
Another key observation in Fig. \ref{fig1} is that, while non-unity 
weights are optimum for small values of $m$, the optimum weights 
approach unity for large $m$. ${\bf G_{pw}}$ being structurally 
similar to ${\bf G_p}$ except for the weights, like ${\bf G_p}$ and 
${\bf G}$ filters, ${\bf G_{pw}}$ is also expected to converge to 
${\bf R}^{-1}$ for $m \rightarrow \infty$, and this explains why
$w_{k,opt}^{(m)} \rightarrow 1$ for $m \rightarrow \infty$.

\vspace{-2mm}
\subsection{Results and Discussion}
\label{sec51}
\vspace{-3mm}
In this subsection, we present a comparison of the bit error rate (BER)
performance of different matrix filters. The various matrix filters
considered include:
$i)$ the conventional filter, ${\bf G}^{(m)}$, given by (\ref{lf2}),
$ii)$ the proposed filter, ${\bf G_p}^{(m)}$, given by (\ref{gp1}),
$iii)$ the MMSE converging filter in \cite{GRL00}, 
	{\small ${\bf G_\mu}^{(m)}$}, given by (\ref{gmu}),
$iv)$ the modified MMSE converging filter, ${\bf G_{p\mu}}^{(m)}$, 
	given by (\ref{mx1}), and
$v)$ the proposed weighted filter, ${\bf G_{pw}}^{(m)}$, given by 
        (\ref{gpw}).

In Fig. \ref{fig2}, we plot the BER performance of the conventional filter, 
${\bf G}^{(m)}$, and the proposed filter, ${\bf G_p}^{(m)}$, as a function 
of the stage index, $m$, for $M=1$, $K=20$, $P=64$, and average SNR = 15 dB, 
for both no near-far (i.e., $A_1=A_2=\cdots=A_K$) as well as near-far 
conditions. In all the simulations, user 1 is taken to be the desired 
user. Random binary sequences are used as spreading sequences. For the 
near-far condition, odd-indexed users (users $3,5,7,\cdots$) transmit
with the same amplitude as the desired user 1, whereas the even-indexed 
users (users $2,4,6,\cdots$) transmit at 10 times larger amplitude than 
the desired user. The performance of the MF detector and the DC detector 
are also plotted for comparison. From Fig. \ref{fig2}, it can be seen 
that the conventional ${\bf G}^{(m)}$ filter approaches the DC detector 
performance rather slowly for increasing $m$. Observe that the performance 
of the proposed ${\bf G_p}^{(m)}$ filter and the conventional ${\bf G}^{(m)}$ 
filter are the same for $m=2$ because of no desired signal loss recovery at 
the 2nd stage of both ${\bf G}^{(m)}$ and ${\bf G_p}^{(m)}$. However, for 
$m\geq3$, the ${\bf G_p}^{(m)}$ filter performs better than the 
${\bf G}^{(m)}$ filter. This is because the ${\bf G_p}^{(m)}$ filter, as 
intended, avoids the generation of new interference and noise terms (e.g., 
{\scriptsize \fbox{$B_I$}} and {\scriptsize \fbox{$B_N$}} terms for $m=3$) 
compared to the ${\bf G}^{(m)}$ filter. The ${\bf G_p}^{(m)}$ filter is 
found to offer greater advantage in near-far conditions, since strong 
other-user interference terms in {\scriptsize \fbox{$B_I$}} are avoided 
in the ${\bf G_p}^{(m)}$.  

Next, in Fig. \ref{fig3}, we present a comparison of the performance of 
the MMSE converging ${\bf G_\mu}^{(m)}$ filter in \cite{GRL00}, and the 
modified MMSE converging, ${\bf G_{p\mu}}^{(m)}$, for the same system 
conditions in Fig. \ref{fig2}. The performance of the MF and MMSE detectors
are also plotted for comparison. Here again, the ${\bf G_\mu}^{(m)}$ and 
${\bf G_{p\mu}}^{(m)}$ filters perform the same for $m=2$. Also, both 
${\bf G_\mu}^{(m)}$ and ${\bf G_{p\mu}}^{(m)}$ are seen to approach the 
MMSE performance as $m$ is increased. For $m\geq 3$, ${\bf G_{p\mu}}^{(m)}$ 
performs better than ${\bf G_\mu}^{(m)}$ because of the avoidance of new 
interference and noise terms. In generating the plot for 
${\bf G_{p\mu}}^{(m)}$, we have used the step sizes in (\ref{eig}), 
which are actually optimum for ${\bf G_\mu}^{(m)}$. Even with these 
step sizes (which can be suboptimum for ${\bf G_{p\mu}}^{(m)}$), the 
proposed ${\bf G_{p\mu}}^{(m)}$ filter approaches the MMSE performance 
faster than the ${\bf G_\mu}^{(m)}$ filter. 

Finally, in Fig. \ref{fig4}, we illustrate the performance of all the
matrix filters considered in this paper, including the proposed
weighted filter, ${\bf G_{pw}}^{(m)}$, under the no near-far condition.
The performance of the MF, DC and MMSE detectors are also plotted. 
It can be observed that among all the filters considered, the proposed 
weighted filter ${\bf G_{pw}}^{(m)}$ performs the best for small values 
of $m$ ($m < 6$, for example). In other words, ${\bf G_{pw}}^{(m)}$
performs best in terms of convergence, i.e., fewer stages are sufficient 
to yield close to DC detector performance. This may be expected, because 
in the ${\bf G_\mu}^{(m)}$ and ${\bf G_{p\mu}}^{(m)}$ filters the optimum 
step sizes are obtained only on a per-stage basis, whereas in the  
${\bf G_{pw}}^{(m)}$ filter the optimum weights are obtained on a 
per-stage as well as a per-user basis. The computation of the optimum 
weights, $w_{k,opt}^{(m)}$, for the ${\bf G_{pw}}^{(m)}$ filter, using 
the closed-form expression in (\ref{opt}), adds to the receiver complexity. 
However, since these optimum weights are computed by using the {\em average}
SINR expression, the weights computation can be carried out off-line once 
(or whenever users exit from or enter into the system, which changes the 
correlation matrix), and this need not add to the per-bit complexity of 
the canceller. 
In Fig. \ref{fig4}, we also show the performance of the conventional 
weighted LPIC given in \cite{TMC06}, denoted by ${\bf G_w}^{(m)}$, for 
up to $m=4$. As can be seen, because of the SIR maximization using optimum
weights, the performance of the ${\bf G_{w}}^{(m)}$ filter is almost the 
same as that of the proposed ${\bf G_{pw}}^{(m)}$ filter. We further note 
that the optimum weights expressions for the ${\bf G_w}^{(m)}$ filter need 
to be derived separately on a stage by stage basis -- the optimum weights 
expressions for up to $m=4$ are given in \cite{TMC06},\cite{mano}, and 
the optimum weights derivation becomes increasingly cumbersome 
for increasing $m$. On the other hand, the feature of making the diagonal
elements zero in the ${\bf G_{pw}}^{(m)}$ filter allows optimum weights
expressions to be obtained for any $m$ (given by Eqn. (\ref{opt})).
In terms of convergence as well as complexity, the 
proposed filter ${\bf G_{p\mu}}^{(m)}$ is also quite attractive.

\vspace{-2mm}
\section{Proposed Filters in Multicarrier DS-CDMA}
\label{sec4}
\vspace{-3mm}
In this section, we extend the proposed matrix filter solutions in the
previous section to multicarrier DS-CDMA (i.e., for $M\geq 2$). Here, 
the multicarrier combining (MCC) operation has to be performed in addition 
to LPIC. Depending on which operation (i.e., MCC or LPIC) gets done 
first at the receiver, the resulting performances and complexities differ. 

{\it {\bfseries Type-I Receiver:}} 
We first consider a receiver where we perform LPIC first on each subcarrier, 
followed by MCC as shown in Fig. \ref{fig5} \cite{icc05}. We refer to this 
receiver as {\em Type-I receiver}. We note that with this Type-I receiver 
using the conventional filter ${\bf G}^{(m)}$, tractable BER analysis 
becomes feasible (reported 
in \cite{icc05}). Also, this receiver architecture can be viewed as a 
direct adoption of the filters proposed for single carrier DS-CDMA in 
Sec. \ref{sec3}, on individual subcarriers in the MC DS-CDMA system. 
Hence, all the matrix filters in Sec. \ref{sec3}, namely, 
${\bf G_p}^{(m)}$, ${\bf G}_{\mu}^{(m)}$, ${\bf G_{p\mu}}^{(m)}$ and 
${\bf G_{pw}}^{(m)}$, can be directly employed on the individual subcarriers. 

{\it {\bfseries Type-II Receiver:}}
Since MCC operation can provide frequency diversity, performing 
MCC before LPIC can enhance the accuracy of the estimates of the MAI 
and hence improve performance. Accordingly, we propose a {\em Type-II
receiver}, where MCC is performed first, followed by LPIC, as shown in 
Fig. \ref{fig7}. The output of the MC combiner in vector form, denoted 
by ${\bf y}^{c(1)}$, can be written as
\begin{eqnarray}
{\bf y}^{c(1)} & = & {\bf R}^{\mbox{\hspace{-0.25mm}}c} \,{\bf b}+{\bf z},
\label{3eq3}
\end{eqnarray}
where
{\small ${\bf y}^{c(1)} = [ \,\, y_{1}^{c(1)}\,\,\,y_{2}^{c(1)}\,\,\,\cdots\,\,\,y_{K}^{c(1)} \,\, ]^T$}, \,
{\small $y_{k}^{c(1)} = \sum_{i=1}^{M} \Big(h_k^{(i)}\Big)^* \, y_k^{(1),(i)}$}, \,
{\small ${\bf R}^c \Define \sum_{i=1}^{M}{{\left({\bf H}^{(i)}\right)}^{\mbox{\hspace{-0.5mm}}H}}{{\bf R}^{(i)}}{{\bf H}^{(i)}}$}, and
{\small ${\bf z}=\mbox{\hspace{-1mm}}\left[\sum_{i=1}^{M}{\left({h_1}^{(i)}\right)}^{*}n_1^{(i)}\,\,\,\, \sum_{i=1}^{M}{\left({h_2}^{(i)}\right)}^*n_2^{(i)}\,\,\cdots\,\,\sum_{i=1}^{M}{\left({h_K}^{(i)}\right)}^*n_K^{(i)}\right]^T$}. The
conventional filter at the MCC output (referred to as ${\bf G}^{c(m)}$
filter) has its $m$th stage output vector given by 
\begin{eqnarray}
& = & \underbrace{\left(\,{\bf I} + \sum_{i=1}^{m-1}\left( {\bf I}-{\bf R_{eff}} \right )^{i} \, \right)}_{{\bf G}^{c(m)}}\,\,\, {\bf y}^{c(1)},
\label{3eq6}
\end{eqnarray}
where ${\bf R_{eff}} \Define {\bf R}^c{\bf H_D}^{-1}$, and
${\bf {H_D}} = \sum_{i=1}^{M}{{\left({\bf H}^{(i)}\right)}^H}{{\bf H}^{(i)}}$.
Similar to the proposed ${\bf G_p}^{(m)}$ filter for single
carrier DS-CDMA in the previous section (where the idea of zeroing the 
diagonal elements of a certain matrix is adopted), a proposed matrix 
filter for MC DS-CDMA Type-II receiver, denoted by ${\bf G_p}^{c(m)}$, 
can be obtained as
\begin{eqnarray}
{\bf y}_{\tiny \bf p}^{c(m)} & = & \underbrace{\left(\sum_{j=0}^{m-1}{\bf B}^c_j\right)}_{{\bf G_p}^{c(m)}} {\bf y}^{c(1)},
\label{p1after}
\end{eqnarray}
where 
${\bf B}^c_n = \Big[{\bf B}^c_{n-1}\left({\bf I}-{\bf R_{eff}} \right)\Big]^{\odot}$,
and ${\bf B}^c_0 = {\bf I}$.

{\it {\bfseries Results and Discussions:}}
In Fig. \ref{fig8}, we present the simulated BER performance comparison
between the Type-I and Type-II receivers for MC DS-CDMA using different
matrix filters. Specifically, we compare the performance of
$i)$ Type-I receiver with the ${\bf G}^{(m)}$ filter,
$ii)$ Type-I receiver with the ${\bf G_p}^{(m)}$ filter,
$iii)$ Type-II receiver with the ${\bf G}^{c(m)}$ filter, and
$iv)$ Type-II receiver with the ${\bf G_p}^{c(m)}$ filter.
We also compare the performance of the above detectors with DC
and MMSE detectors. Random binary sequences of length $P$ are used 
as the spreading sequences on each subcarrier, and the average SNR 
of user $k$ is defined as $\frac{MA_k^2}{\sigma^2}$.
In Fig. \ref{fig8}, we plot the BER 
as a function of stage index, $m$, for $M=4$, $K=20$, $P=64$, and 
average SNR = 14 dB. The following observations can be made from 
Fig. \ref{fig8}:
\vspace{-5mm}
\begin {itemize}
\item Comparing the performance of Type-I and Type-II receivers
for a given filter, we observe that Type-II receivers perform 
significantly better than Type-I receivers. For example, comparing the 
performance of Type-I receiver with the ${\bf G}^{(m)}$ filter and Type-II 
receiver with the ${\bf G}^{c(m)}$ filter, we see that Type-II receiver with 
the ${\bf G}^{c(m)}$ filter performs significantly better (e.g., for $m=4$, 
Type-I with the ${\bf G}^{(m)}$ filter gives a BER of $8 \times 10^{-2}$, 
whereas Type-II with the ${\bf G}^{c(m)}$ filter results in a BER of 
$2 \times 10^{-4}$). The superiority of Type-II receivers is consistent 
across all filters considered, i.e., ${\bf G}$, ${\bf G_p}$, DC, MMSE. 
This superiority of Type-II receivers is expected, since the 
MAI estimates can be more accurate in Type-II, because of multicarrier 
combining before IC. 
\vspace{-2mm}
\item Like in SC DS-CDMA, in MC DS-CDMA, the proposed 
${\bf G_p}$ filter performs better than the ${\bf G}$ 
filter. This is observed to be true in both Type-I as well as Type-II 
receivers. For example, Type-I with the ${\bf G_p}^{(m)}$ filter
achieves a BER of $9 \times 10^{-3}$ in just 5 stages, whereas the same 
BER is achieved by Type-I with the ${\bf G}^{(m)}$ filter only after 15 
stages. Likewise, in Type-II receivers, the ${\bf G_p}^{c(m)}$ filter 
performs better than the ${\bf G}^{c(m)}$ filter.
\end {itemize}

\vspace{-9mm}
\section{Conclusions}
\label{sec5}
\vspace{-7mm}
We proposed improved LPICs for CDMA by viewing an LPIC as a linear matrix 
filter. Specifically, we proposed new linear matrix filters which achieved
better performance than other linear matrix filters in the literature. This 
was made possible by avoiding the generation of certain new interference 
and noise terms by making the diagonal elements of a certain matrix 
equal to zero 
in each stage, without adding complexity. We also extended the proposed 
matrix filter solutions to multicarrier DS-CDMA, where we considered two 
types of receivers; in both types of receivers the proposed filters were 
shown to outperform other filters in the literature.  

\vspace{-3mm}
\subsection*{Appendix A: \hspace{1mm} {\normalsize Expression for the
3rd stage ${\bf G}^{(m)}$ filter output}}
\vspace{-3mm}

In this appendix, we write the expression for the 3rd stage output
of the ${\bf G}^{(m)}$ filter (i.e., CLPIC) in an expanded form. 
From (\ref{lf2}), ${\bf y}^{(3)}$ can be written as
\begin{eqnarray}
{\bf y}^{(3)} & \mbox{\hspace{-4mm}} = \mbox{\hspace{-2mm}} & \mbox{\hspace{-2mm}} 
\left[{\bf I}+({\bf I}-{\bf R})+({\bf I}-{\bf R})^2\right]{\bf y}^{(1)} 
\,\, = \,\,
{\bf y}^{(2)} + ({\bf I}-{\bf R})^{2} \, {\bf y}^{(1)} \nonumber \\
& \mbox{\hspace{-2mm}} = &  \mbox{\hspace{-0mm}} 
y_k^{(2)} + \sum_{j=1}^{K} \,\, \sum_{i\ne k,j}^{K}\rho_{ki}\rho_{ij}y_j^{(1)} 
\,\, = \,\,\, y_k^{(2)} + \,\, \underbrace{\sum_{i\ne k}^{K}\rho_{ki}\rho_{ik}y_k^{(1)}}_{T_1: \,\, \mbox{\footnotesize{case of}} j=k}+\underbrace{\sum_{j\ne k}^{K} \,\, \sum_{i\ne k,j}^{K}\rho_{ki}\rho_{ij}y_j^{(1)}}_{T_2: \,\, \mbox{\footnotesize{case of}} \,j \neq k}. 
\label{eqn3a}
\end{eqnarray}
We point out that the term $T_1$ in the above equation recovers the
desired signal lost in the 2nd stage, and the term $T_2$ removes the
interference terms generated in the 2nd stage. Substituting (\ref{A2})
and (\ref{3eq1}) in (\ref{eqn3a}), we can write

\vspace{-9mm}
\begin{eqnarray}
y_k^{(3)} & \mbox{\hspace{-4mm}} = \mbox{\hspace{-2mm}} & \mbox{\hspace{-2mm}} 
x_k\left(1-\sum_{j\neq k}^{K}\rho_{kj}\rho_{jk}\right) - \sum_{j\neq k}^{K} \,\, \sum_{l\neq j,k}^{K}\rho_{kj}\rho_{jl} x_l + n_k - \sum_{j\neq k}^{K}\rho_{kj} n_j \nonumber \\
& & + \,\, \sum_{i\ne k}^{K}\rho_{ki}\rho_{ik}\left[x_k+\sum_{j\ne k}^{K}\rho_{kj}x_j+n_k\right] 
+ \,\, \sum_{j\ne k}^{K} \,\, \sum_{i\neq k,j}^{K}\rho_{ki}\rho_{ij}\left[x_j+\sum_{l\ne j}^{K}\rho_{jl}x_l+n_j\right] \nonumber \\
& \mbox{\hspace{-2mm}} = &  \mbox{\hspace{-2mm}} 
x_k - \sum_{j\neq k}^{K}\rho_{kj}\rho_{jk}x_k - \sum_{j\neq k}^{K}\sum_{l\neq j,k}^{K}\rho_{kj}\rho_{jl} x_l+n_k-\sum_{j\neq k}^{K}\rho_{kj} n_j\nonumber\\
& & + \,\, \sum_{i\ne k}^{K}\rho_{ki}\rho_{ik}x_k+\sum_{i\ne k}^{K}\rho_{ki}\rho_{ik}\sum_{j\ne k}^{K}\rho_{kj}x_j+\sum_{i\ne k}^{K}\rho_{ki}\rho_{ik}n_k\nonumber\\
& & + \,\, \sum_{j\ne k}^{K}\sum_{i\neq k,j}^{K}\rho_{ki}\rho_{ij}x_j + \sum_{j\ne k}^{K} \,\, \sum_{i\neq k,j}^{K}\rho_{ki}\rho_{ij} 
\sum_{l\ne j}^{K}
\rho_{jl}x_l 
+ \,\, \sum_{j\ne k}^{K}\sum_{i\neq k,j}^{K}\rho_{ki}\rho_{ij}n_j \nonumber \\
& & \mbox{\hspace{-24mm}} = \,\, x_k - \underbrace{\sum_{j\neq k}^{K}\rho_{kj}\rho_{jk}x_k}_{{\tiny {\fbox A}}}-\underbrace{\sum_{j\neq k}^{K}\sum_{l\neq j,k}^{K}\rho_{kj}\rho_{jl}x_l}_{{\tiny {\fbox C}}} + \,\, n_k-\sum_{j\neq k}^{K}\rho_{kj}n_j\nonumber\\
& & \mbox{\hspace{-24mm}} + \,\, \underbrace{\sum_{i\ne k}^{K}\rho_{ki}\rho_{ik}x_k}_{{\tiny {\fbox A}}}+\underbrace{\sum_{i\ne k}^{K}\rho_{ki}\rho_{ik}\sum_{j\ne k}^{K}\rho_{kj}x_j}_{{\tiny {\fbox {$B_I$}}}}+\underbrace{\sum_{i\ne k}^{K}\rho_{ki}\rho_{ik}n_k}_{{\tiny \mbox{\fbox{$B_N$}}}}\nonumber\\
& & \mbox{\hspace{-24mm}} + \,\, \underbrace{\sum_{j\ne k}^{K} \,\, \sum_{i\neq k,j}^{K}\rho_{ki}\rho_{ij}x_j}_{{\tiny {\fbox C}}}+\underbrace{\sum_{j\ne k}^{K} \,\, \sum_{i\neq k,j}^{K}\rho_{ki}\rho_{ij}\rho_{jk}x_k}_{{\tiny{ \fbox D}}} 
+ \,\, \underbrace{\sum_{j\ne k}^{K} \,\, \sum_{i\neq k,j}^{K}\rho_{ki}\rho_{ij}\sum_{l\ne k,j}^{K}\rho_{jl}x_l}_{{\tiny {\fbox {$E_I$}}}}+\underbrace{\sum_{j\ne k}^{K}\sum_{i\neq k,j}^{K}\rho_{ki}\rho_{ij}n_j}_{{\tiny{ \fbox {$E_N$}}}}.
\label{3rd}
\end{eqnarray}

\vspace{-3mm}
\subsection*{Appendix B: \hspace{1mm} {\normalsize Expression for the $m$th stage ${\bf G}^{(m)}$ filter output, $m > 3$}}
\vspace{-3mm}

From (\ref{lf2}), we can write $y_k^{(m)}$ for $m\geq3$ as
\begin{eqnarray}
y_k^{(m)} & = & y_k^{(1)}-\left(\sum_{k_1\neq k}^{K}\rho_{kk_1}-\sum_{k_1=1}^{K} \left(\sum_{k_2\ne k,k_1}^{K}\rho_{kk_2}\rho_{k_2k_1}-\sum_{k_2\ne k_1}^{K}\sum_{k_3\ne k,k_2}^{K}\rho_{kk_3}\rho_{k_3k_2}\rho_{k_2k_1}+\cdots \right.\right. \nonumber\\
& & \mbox{\hspace{-14mm}}\left.\left.+ \, (-1)^{m-1}\hspace{-2mm}\sum_{k_2\ne k_1}^{K}\hspace{-0mm}\sum_{k_3\ne k_2}^{K}\cdots\sum_{k_{m-2}\ne k_{m-3}}^{K}\sum_{k_{m-1}\ne k,k_{m-2}}^{K}\hspace{-7mm}\rho_{kk_{m-1}}\rho_{k_{m-1}k_{m-2}}\cdots \rho_{k_3k_2}\rho_{k_2k_1}\right)\right)y_{k_1}^{(1)}.
\label{3eq71}
\end{eqnarray}
Equivalently, (\ref{3eq71}) can be written as 
\begin{eqnarray}
y_k^{(m)} & = & y_k^{(m-1)}+(-1)^{m+1}\hspace{-0mm}\sum_{k_1=1}^{K}\sum_{k_2\ne k_1}^{K}\hspace{-0mm}\sum_{k_3\ne k_2}^{K}\cdots\hspace{-2mm}\sum_{k_{m-2}\ne k_{m-3}}^{K} \nonumber \\
& & \sum_{k_{m-1}\ne k,k_{m-2}}^{K}\hspace{-5mm}\rho_{kk_{m-1}}\rho_{k_{m-1}k_{m-2}}\cdots\rho_{k_3k_2}\rho_{k_2k_1} \,y_{k_1}^{(1)}.
\label{3eq81}
\end{eqnarray}
The first summation in the 2nd term on the RHS of (\ref{3eq81}) can be
split into two terms, one for $k_1=k$ and another for $k_1\neq k$, as
\begin{eqnarray}
y_k^{(m)} & = & y_k^{(m-1)}+\underbrace{(-1)^{m+1}\hspace{-0mm}\sum_{k_2\ne k}^{K}\hspace{-0mm}\sum_{k_3\ne k_2}^{K}\cdots\sum_{k_{m-2}\ne k_{m-3}}^{K}\sum_{k_{m-1}\ne k,k_{m-2}}^{K}\hspace{-5mm}\rho_{kk_{m-1}}\rho_{k_{m-1}k_{m-2}}\cdots \rho_{k_3k_2}\rho_{k_2k} \, y_{k}^{(1)}}_{T_3}\nonumber\\
& & \mbox{\hspace{-10mm}}+ \, \underbrace{(-1)^{m+1}\hspace{-0mm}\sum_{k_1\ne k}^{K}\sum_{k_2\ne k_1}^{K}\hspace{-0mm}\sum_{k_3\ne k_2}^{K}\cdots\sum_{k_{m-2}\ne k_{m-3}}^{K}\sum_{k_{m-1}\ne k,k_{m-2}}^{K}\hspace{-5mm}\rho_{kk_{m-1}}\rho_{k_{m-1}k_{m-2}}\cdots \rho_{k_3k_2}\rho_{k_2k_1} \, y_{k_1}^{(1)}}_{T_4}.
\label{3eq91}
\end{eqnarray}
For $m\geq4$, $y_k^{(m-1)}$ can be written in an alternate form as
\begin{eqnarray}
y_k^{(m-1)}& =& x_k + \underbrace{(-1)^{m}\hspace{-0mm}\sum_{k_2\ne k}^{K}\hspace{-0mm}\sum_{k_3\ne k_2}^{K}\cdots\sum_{k_{m-2}\ne k_{m-3}}^{K}\sum_{k_{m-1}\ne k,k_{m-2}}^{K}\hspace{-5mm}\rho_{kk_{m-1}}\rho_{k_{m-1}k_{m-2}}\cdots \rho_{k_3k_2}\rho_{k_2k} x_k}_{T_5: \,\,\,\, {\footnotesize \mbox{desired signal loss/gain at the (m-1)th stage output}}}\nonumber\\
& &\mbox{\hspace{-15mm}} + \, \underbrace{(-1)^{m}\hspace{-0mm}\sum_{k_1\ne k}^{K}\sum_{k_2\ne k_1}^{K}\hspace{-0mm}\sum_{k_3\ne k_2}^{K}\cdots\sum_{k_{m-2}\ne k_{m-3}}^{K}\sum_{k_{m-1}\ne k,k_{m-2}}^{K}\hspace{-5mm}\rho_{kk_{m-1}}\rho_{k_{m-1}k_{m-2}}\cdots \rho_{k_3k_2}\rho_{k_2k_1} x_{k_1}}_{T_6: \,\,\,\, {\footnotesize \mbox{new interference terms generated at the (m-1)th stage output}}}\nonumber\\
& & \mbox{\hspace{-15.0mm}} + \, n_k - \sum_{k_1\ne k}^{K}\rho_{kk_1}n_{k_1}+\sum_{k_1=1}^{K}\sum_{k_2\ne k,k_1}\rho_{kk_2}\rho_{k_2k_1}n_{k_1}- \cdots \nonumber\\
& & \mbox{\hspace{-15mm}} + \, (-1)^{m}\hspace{-0mm}\sum_{k_1=1}^{K}\sum_{k_2\ne k_1}^{K}\hspace{-0mm}\sum_{k_3\ne k_2}^{K}\cdots\sum_{k_{m-3}\ne k_{m-4}}^{K}\sum_{k_{m-2}\ne k,k_{m-3}}^{K}\hspace{-5mm}\rho_{kk_{m-2}}\rho_{k_{m-2}k_{m-3}}\cdots \rho_{k_3k_2}\rho_{k_2k_1}n_{k_1}.
\label{3eq101}
\end{eqnarray}
\baselineskip 1.70pc
Comparing the output terms in stages $m-1$ and $m$ in Eqns. (\ref{3eq101})
and (\ref{3eq91}), respectively, we can observe the following.
\vspace{-4mm}
\begin{enumerate}
\item 	The desired signal loss/gain that occurred in the $(m-1)$th stage 
	(i.e., $T_5$ in (\ref{3eq101})) is recovered/removed in the $m$th 
	stage (see $T_3$ in (\ref{3eq91}) and note that $y_k^{(1)}$ in it 
	has $x_k$). In the process, new interference terms proportional 
	to $\rho^m$ and additional noise terms proportional to $\rho^{m-1}$ 
	get generated, due to all terms other than $x_k$ in $y_k^{(1)}$ in 
	$T_3$. Note that for the case of $m=3$, these interference and 
	noise terms generated are given by {\tiny \fbox{$B_I$}} and
	{\tiny \fbox{$B_N$}} in Eqn. (\ref{3rd}) in Appendix A.
\vspace{-2mm}
\item 	The new interference terms that were generated in the $(m-1)$th 
	stage (i.e., $T_6$ in (\ref{3eq101})) are removed in the $m$th 
	stage (see $T_4$ in (\ref{3eq91}) and note that $y_{k_1}^{(1)}$ 
	in it has $x_{k_1}$). In the process, new interference terms 
	proportional to $\rho^m$ and additional noise terms proportional 
	to $\rho^{m-1}$ get generated, due to all terms other than 
	$x_{k_1}$ in $y_{k_1}^{(1)}$ in $T_4$. Note that for the case 
	of $m=3$, these interference and noise terms are given by 
	{\tiny \fbox{$E_I$}} and {\tiny \fbox{$E_N$}} in Eqn. (\ref{3rd}) 
	in Appendix A.
\end{enumerate}

\vspace{-8mm}
{\footnotesize 
\bibliographystyle{IEEE}

}

\newpage 

\begin{figure}
\begin{center}
\epsfxsize=10.0cm
\epsfbox{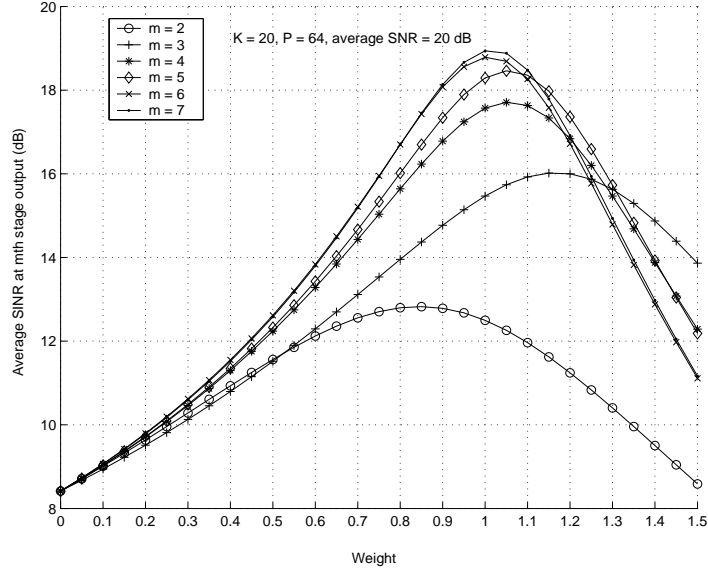}
\caption{Average output SINR as a function of weight $w_k^{(m)}$ for
the proposed ${\bf G_{pw}}^{(m)}$ filter. $M=1$ (i.e., single carrier
DS-CDMA), $K=20$, $P=64$, average SNR = 20 dB. No near-far condition.} 
\label{fig1}
\end{center}
\end{figure}

\begin{figure}
\begin{center}
\epsfxsize=10.0cm
\epsfbox{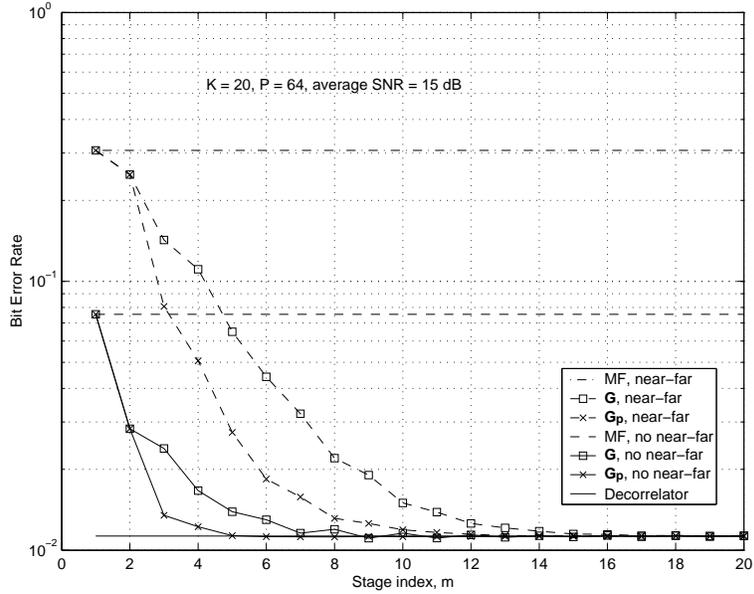}
\caption{BER performance of various linear matrix filters --
$i)$ conventional filter ${\bf G}^{(m)}$, and $ii)$ proposed filter
${\bf G_p}^{(m)}$. $M=1$, $K=20$, $P=64$, average SNR = 15 dB. 
Near-far as well as no near-far conditions. }
\label{fig2}
\end{center}
\end{figure}

\begin{figure}
\begin{center}
\epsfxsize=10.0cm
\epsfbox{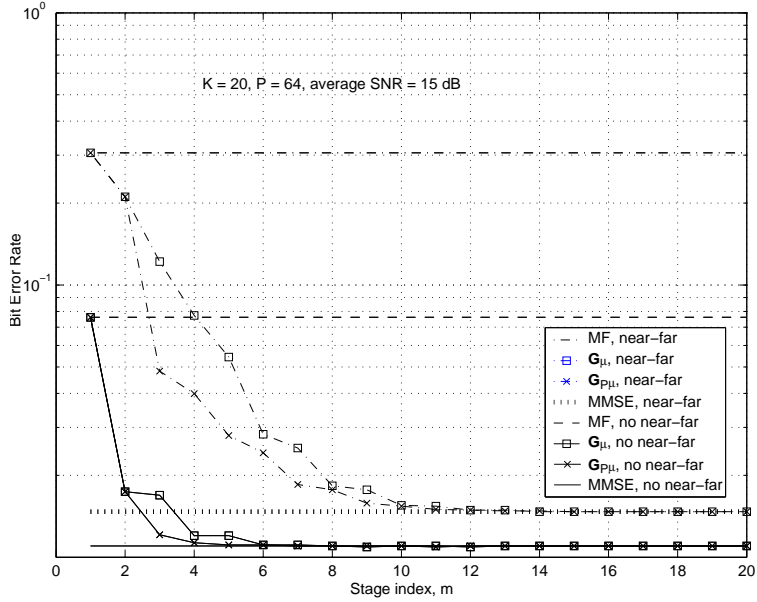}
\caption{BER performance of various linear matrix filters --
$i)$ MMSE converging filter ${\bf G_\mu}^{(m)}$, and
$ii)$ modified MMSE converging filter, ${\bf G_{p\mu}}^{(m)}$.
$M=1$, $K=20$, $P = 64$, average SNR = 15 dB. Near-far as 
well as no near-far conditions. }
\label{fig3}
\end{center}
\end{figure}

\begin{figure}
\begin{center}
\epsfxsize=10.0cm
\epsfbox{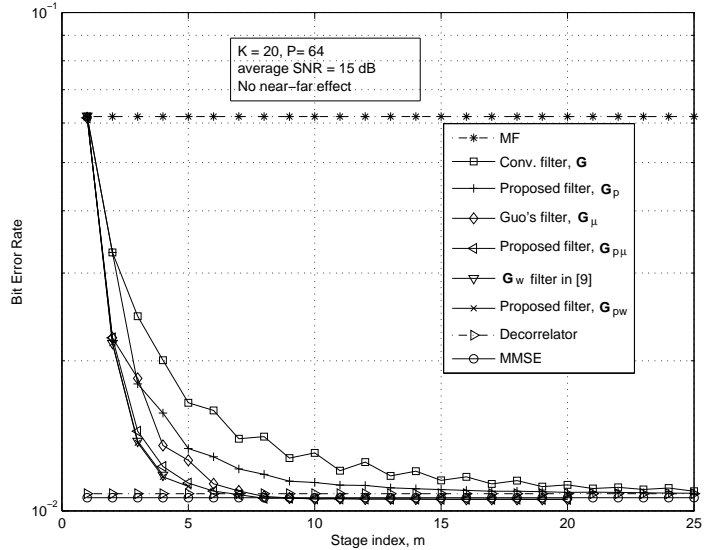}
\caption{BER performance of various linear matrix filters --
$i)$ ${\bf G}^{(m)}$ filter, $ii)$ proposed ${\bf G_p}^{(m)}$ filter, 
$iii)$ MMSE converging ${\bf G_\mu}^{(m)}$ filter, $iv)$ modified
MMSE converging filter, ${\bf G_{p\mu}}^{(m)}$, $v)$ proposed 
weighted filter, ${\bf G_{pw}}^{(m)}$, and $vi)$ conventional weighted 
LPIC filter in \cite{TMC06}, ${\bf G_w}^{(m)}$. $M=1$, $K=20$, $P = 64$, 
average SNR = 15 dB. No near-far condition.}
\label{fig4}
\end{center}
\end{figure}

\begin{figure}
\begin{center}
\epsfxsize=9.5cm
\epsfbox{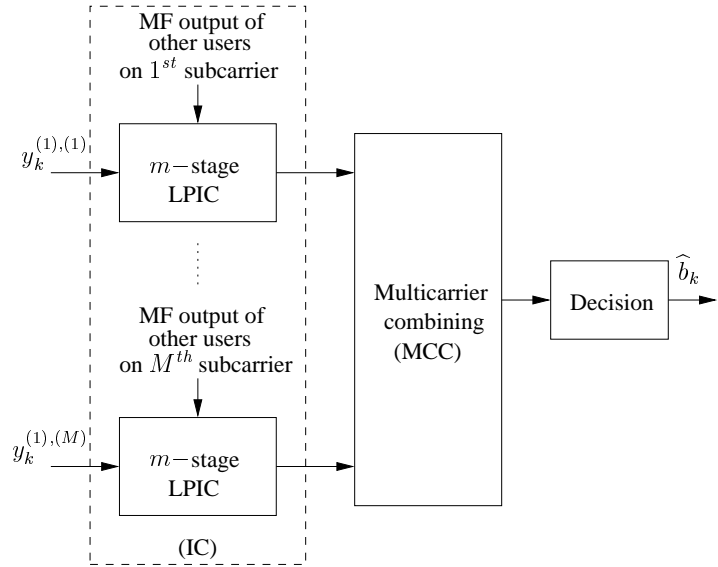}
\caption{Type-I Receiver for multicarrier DS-CDMA, $M\geq 2$. LPIC is 
done first on each subcarrier. Multicarrier Combining is done next.  }
\label{fig5}
\end{center}
\end{figure}

\begin{figure}
\begin{center}
\epsfxsize=12.0cm
\epsfbox{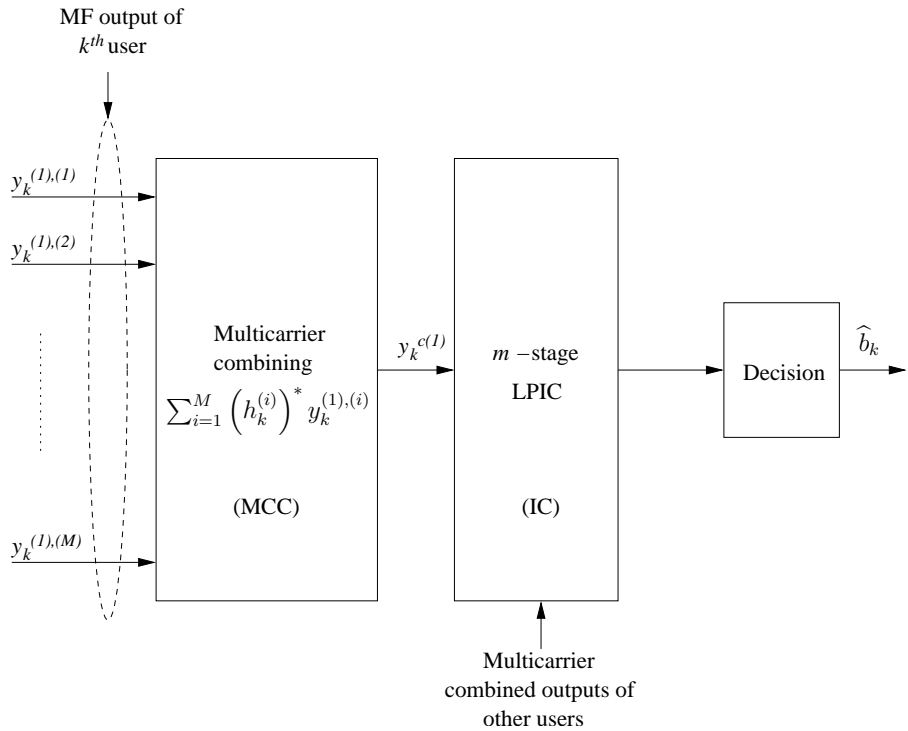}
\caption{Type-II Receiver for MC DS-CDMA, $M\geq 2$. LPIC is done after 
multicarrier combining.}
\label{fig7}
\end{center}
\end{figure}

\begin{figure}
\begin{center}
\epsfxsize=10cm
\epsfbox{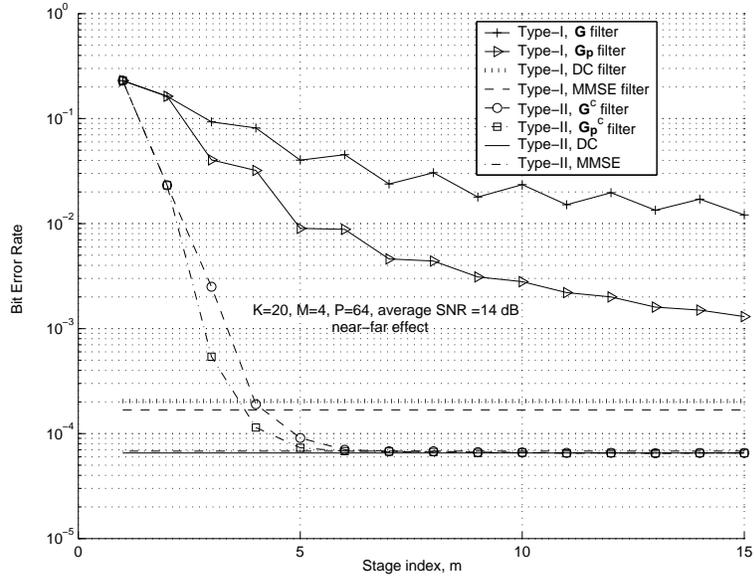}
\caption{BER performance of the proposed Type-I and Type-II receivers as 
a function of stage index $m$ for MC DS-CDMA with -- 
$i)$ Type-I with ${\bf G}^{(m)}$ filter, 
$ii)$ Type-I with ${\bf G_p}^{(m)}$ filter, 
$iii)$ Type-II with ${\bf G}^{c(m)}$ filter, and 
$iv)$ Type-II with ${\bf G_p}^{c(m)}$ filter.
$M=4$, $K=20$, $P=64$, average SNR = 14 dB. Near-far effect
as described in Sec. \ref{sec51}. }
\label{fig8}
\end{center}
\end{figure}

\end{document}